\documentclass[MSNbibl,nameyear,dvips]{arxstspdf}
\usepackage{flushend}
\usepackage{stfloats}
\volume{26}
\issue{3}
\pubyear{2011}
\firstpage{369}
\lastpage{387}
\doi{10.1214/11-STS358}

\begin{document}
\begin{frontmatter}

\title{Covariance Estimation: The GLM and Regularization Perspectives}

\runtitle{Modeling Covariance Matrices}

\begin{aug}
\author{\fnms{Mohsen} \snm{Pourahmadi}\corref{}\ead[label=e1]{pourahm@stat.tamu.edu}}

\runauthor{M. Pourahmadi}

\affiliation{Texas A\&M University}

\address{Mohsen Pourahmadi is Professor, Department of Statistics,
Texas A\&M University, College Station, Texas 77843-3143, USA
\printead{e1}.}

\end{aug}

\begin{abstract}
Finding an \textit{unconstrained} and \textit{statistically
interpretable} re\-parameterization of a covariance matrix is still an
open problem in statistics. Its solution is of central importance in
covariance estimation, particularly in the recent high-dimensional data
environment where enforcing the positive-definiteness constraint could
be computationally expensive. We provide a survey of the progress made
in modeling covariance matrices from two relatively complementary
perspectives: (1)~generalized linear models (GLM) or parsimony and use
of covariates in low dimensions, and (2) regularization or sparsity for
high-dimensional data. An emerging, unifying and powerful trend in both
perspectives is that of reducing a covariance estimation problem to
that of estimating a sequence of regression problems. We point out
several instances of the regression-based formulation. A notable case
is in sparse estimation of a precision matrix or a Gaussian graphical
model leading to the fast graphical LASSO algorithm. Some advantages
and limitations of the regression-based Cholesky decomposition relative
to the classical spectral (eigenvalue) and variance-correlation
decompositions are highlighted. The former provides an unconstrained
and statistically interpretable reparameterization, and guarantees the
positive-definiteness of the estimated covariance matrix. It reduces
the unintuitive task of covariance estimation to that of modeling a
sequence of regressions at the cost of imposing an \textit{a priori
order} among the variables. Elementwise regularization of the sample
covariance matrix such as banding, tapering and thresholding has
desirable asymptotic properties and the sparse estimated covariance
matrix is positive definite with probability tending to one for large
samples and dimensions.
\end{abstract}


\end{frontmatter}

\section{Introduction} \label{intro}

The $p \times p$ covariance matrix $\Sigma$ of a random vector
$Y=(y_1,\ldots,y_p)^\prime$
 with as many as $\frac{p(p+1)}{2}$
constrained parameters plays a central role in virtually all of
classical multivariate statistics (Anderson, \citeyear{And03}), time
series analysis (Box, Jenkins and Reinsel, \citeyear{BoxJenRei94}),
spatial data analysis (Cressie, \citeyear{Cre91}), variance components
and longitudinal data analysis (Searle, Casella and McCulloch,
\citeyear{SeaCasMcC92}; Diggle et al., \citeyear{Digetal02}), and in
the modern and rapidly growing area of statistical and machine learning
dealing with massive and high-dimensional data (Hastie, Tibshirani and
Friedman, \citeyear{HasTibFri09}). It is generally recognized that the
two major challenges in covariance estimation are the
positive-definiteness constraint and the high-dimensionality where the
number of parameters grows quadratically in $p$. In this survey, we
point out that these challenges become manageable, for example, by
reducing covariance estimation to that of solving a series of
(penalized) least squares regression problems.

$\!\!$Nowadays, in microarray data, spectroscopy, finan\-ce, climate studies
and abundance data in community ecology it is common to have situations
where $p\gg n $. Here the use of a sample covariance matrix is
problematic (Stein, \citeyear{Ste56}), particularly when its inverse is
needed as, for example, in classification procedures (Anderson,
\citeyear{And03}, Chapter 6), multivariate linear regression (Warton,
\citeyear{War08}; Witten and Tibshirani, \citeyear{WitTib09}),
portfolio selection (Ledoit, Santa-Clara and Wolf,
\citeyear{LedSanWol03}) and Gaussian graphical models (Wong, Carter and
Kohn, \citeyear{WonCarKoh03}; Meinshausen and B\"{u}hlmann,
\citeyear{MeiBuh06}; Yuan and Lin, \citeyear{YuaLin07}). In these
situations and others, it is desirable to find alternative covariance
estimators that are more accurate and better-conditioned than the
sample covariance matrix.

It was noted rather early by Stein (\citeyear{Ste56,Ste75}) that the
sample covariance matrix $S=\frac {1}{n}\sum_{i=1}^n Y_iY_i^\prime,$
based on a sample of size $n$ from a mean zero normal population with
the covariance matrix $\Sigma$, though unbiased and positive definite,
is a poor estimator when $\frac{p}{n}$ is large (Johnstone,
\citeyear{Joh01}). It distorts the eigenstructure of $\Sigma$, in the
sense that the largest (smallest) sample eigenvalue will be biased
upward (downward). Since then many improved estimators have been
proposed by shrinking the eigenvalues of~$S$ toward a central value
(Haff, \citeyear{Haf80,Haf91}; Lin and Perlman, \citeyear{LinPer85};
Dey and Srinivasan, \citeyear{DeySri85}; Yang and Berger,
\citeyear{YanBer94}; Ledoit and Wolf, \citeyear{LedWol04}). These have
been derived from a decision-theoretic perspective or by specifying an
appropriate prior for the covariance matrix. The Stein's family of
shrinkage estimators leaving intact the eigenvectors of the sample
covariance matrix are neither sparse nor parsimonious. However, lately
the search for sparsity and parsimony has led to either shrinking the
matrix~$S$ itself toward certain targets like diagonal and
autoregressive structures as in Daniels and Kass
(\citeyear{DanKas99,DanKas01}), or shrinking its eigenvectors as in
Hoff (\citeyear{Hof09}) and Johnstone and Lu (\citeyear{JohLu09}).

In many applications the need for the precision matrix $\Sigma^{-1}$ is
stronger than that for $\Sigma$ itself. Though the former can be
computed from the latter in ${\mathcal O}(p^3)$ operations, this could
be computationally expensive and should be avoided when $p$ is large.
The regression-based approach of Meinshausen and  B\"{u}hl\-mann
(\citeyear{MeiBuh06}) provides a sparse estimate of the precision
matrix or a Gaussian graphical model by fitting separate LASSO
regression to each variable, using the others as predictors. This
simple idea has inspired several direct and improved sparse estimators
of $\Sigma^{-1}$ using a penalized likelihood approach with a~LASSO
penalty on its off-diagonal terms (Yuan and Lin, \citeyear{YuaLin07};
Banerjee, El~Ghaoui and d'Aspremont, \citeyear{BanElGdAs08}; Friedman,
Hastie and Tibshirani, \citeyear{FriHasTib08}; Rothman et al.,
\citeyear{Rotetal08}; Rocha, Zhao and Yu, \citeyear{RocZhaYu}; Peng et
al., \citeyear{PenZhoZhu09}). Friedman, Hastie and Tibshirani
(\citeyear{FriHasTib08}) graphical LASSO is the fastest available
algorithm to date.  Surprisingly, such a sparse covariance estimator is
guaranteed to be positive definite (Banerjee, El~Ghaoui and
d'Aspremont, \citeyear{BanElGdAs08}).

A remarkable unifying \textit{regression-based} theme has emerged from
research on covariance estimation in the last decade or so. Some
notable examples are as follows: (i) formulating principal component
analysis (PCA) as regression optimization problems (Jong and Kotz,
\citeyear{JonKot99}; Zou, Hastie and Tibshirani,
\citeyear{ZouHasTib06}), sparse loadings are then estimated by imposing
the lasso constraint on the regression coefficients, (ii)
regression-based derivation and interpretation of the modified Cholesky
decomposition of a covariance matrix and its inverse (Pourahmadi,
\citeyear{Pou99,Pou01}, Section~3.5; Bilmes, \citeyear{Bil00}; Huang et
al., \citeyear{Huaetal06}; Rothman, Levina and Zhu,
\citeyear{RotLevZhu10}), (iii) the regression approach of Meinshausen
and  B\"{u}hlmann (\citeyear{MeiBuh06}) to the Gaussian graphical
models, (iv) the graphical LASSO algorithm of Friedman, Hastie and
Tibshirani (\citeyear{FriHasTib08,FriHasTib}) and (v) the iteratively
reweighted penalized likelihood of Fan, Feng and Wu
(\citeyear{FanFenWu09}) where nonconcave penalties such as the smoothly
clipped absolute deviation (SCAD) are imposed on the entries of the
precision matrix. The problem of sparse estimation of the precision
matrix is then recast as a sequence of penalized likelihood problems
with a weighted LASSO penalty and solved using the graphical LASSO
algorithm of Friedman, Hastie and Tibshirani (\citeyear{FriHasTib08}).

Among these approaches it seems only (ii) has the expressed goal of
providing unconstrained and statistically interpretable regression
parameters for the covariance\vadjust{\goodbreak} (precision) matrix. Unfortunately,
however, unlike the others which work for unordered variables and
provide permutation-invariant covariance estimators, (ii) and a few
other alternatives to the sample covariance matrix proposed in recent
years give rise to covariance estimators which are sensitive to the
order among the variables in $Y$. These approaches are suitable for
time series and longitudinal data which have a natural (time) order
among the variables in $Y$, and assume that variables far apart in the
ordering are less correlated. For example, regularizing a covariance
matrix by tapering (Furrer and Bengtsson, \citeyear{FurBen07}), banding
(Bickel and Levina, \citeyear{BicLev04,BicLev08N1}; Wu and Pourahmadi,
\citeyear{WuPou03,WuPou09}) and generally those based on the Cholesky~de\-composition
of the covariance matrix or its inverse (Pourahmadi,
\citeyear{Pou99,Pou00}; Rothman, Levina and Zhu,
\citeyear{RotLevZhu10}) do impose an order among the components of~$Y$
and are not permutation-invariant. The idea of thresholding individual
entries of $S$ has been used in the estimation of large covariance
matrices by Bickel and Levina (\citeyear{BicLev08N2}), El Karoui
(\citeyear{ElK08N1,ElK08N2}) and Rothman, Levina and Zhu
(\citeyear{RotLevZhu09}). Such estimators are permutation-invariant
with desirable asymptotic properties.

It should be noted that the recent surge of interest in
regression-based approaches to \textit{sparsity} in high-dimensional
data bodes well with the long history of interest in \textit{parsimony}
and using covariates when modeling covariance matrices of
low-dimensional da\-ta (Anderson, \citeyear{And73}). For example,
longitudinal data collected from expensive clinical trials and
biological experiments may have about $n=30$ subjects and $p \le 10$
measurements per subject. Parsimonious and accurate modeling of the
covariance structure is important in these application areas (Cannon et
al., \citeyear{Canetal01}; Carroll, \citeyear{Car03};  Fitzmaurice et al., \citeyear{autokey42}). However,
the area of data-based covariance modeling is woefully underdeveloped.
At present, a practitioner has the option of picking a structured
covariance matrix from a long menu, where at one extreme the choice is
$\sigma^2I_p$ (independence) and at the other the unstructured
covariance matrix with $\frac {p(p+1)}{2}$ parameters (Zimmerman and
N\'{u}\~{n}ez-Ant\'{o}n, \citeyear{ZimNun01,ZimNun10}). Of course, it
is desirable to bridge the gap between these two extremes and develop a
bona fide GLM methodology and a data-based framework for modeling
covariance matrices. Attempts to develop such methods going beyond the
traditional linear covariance models (Anderson, \citeyear{And73}) have
been made in recent years by Chiu, Leonard and Tsui
(\citeyear{ChiLeoTsu96}) and Pourahmadi (\citeyear{Pou99,Pou00}); Pan
and MacKenzie\break (\citeyear{PanMac03});
Lin and Wang (\citeyear{LinWan09}); Leng, Zhang and Pan
(\citeyear{LenZhaPan10}); Lin (\citeyear{Lin}) using the spectral and
Cholesky decompositions of covariance matrices, respectively.

Given the complex nature of the positive-definite\-ness constraint, in
 developing  a GLM methodolgy it is plausible
to factorize $\Sigma$ into two or more components capturing the
``variance'' through a diagonal matrix and the ``dependence'' through a
matrix with $\frac {p(p-1)}{2}$ functionally unrelated entries. A
decomposition is ideal for the GLM purposes, if its ``dependence''
component is an unconstrained and statistically interpretable matrix.
 The three most commonly used decompositions in
increasing order of adherence to the GLM principles are the
variance-correlation, spectral and Cholesky decompositions where their
``dependence'' components are correlation, orthogonal and lower
triangular matrices, respectively. While the entries of the first two
matrices are always constrained, those of the last are unconstrained.
Interestingly, these three decompositions are subsumed (Zimmerman and
N\'{u}\~{n}ez-Ant\'{o}n, \citeyear{ZimNun01}, page 59) by
 a decomposition from the class of factor/mixed models (Anderson, \citeyear{And03}):
\begin{equation}\label{factor}
\Sigma=ZBZ'+W.
\end{equation}
Here, the matrix $Z$ is $p\times q$ with $q$ standing for the number of
latent factors, $B$ and $W$ are $q\times q$ and $p\times p$ unknown
preferably diagonal matrices. The representation (\ref{factor}) is
valid only when each of the $p$ variables are well-approximated  as
linear combinations of the same latent factors plus an independent
error. In principle, this may occur when $q$ is large, and adding $W$
to the reduced rank decomposition ensures the positive-definiteness of
$\Sigma$. Technical difficulties with the use of (\ref{factor}) can be
resolved to various extents by choosing the components of the quadruple
$(q,W,B,Z)$ close to the ideal values of $q=p$, $W=0$, $B$ diagonal and
$Z$ sparse or structured.

The outline of the paper is as follows. Section~\ref{sec:overview}
covers some preliminaries on the GLM for covariance matrices, the three
standard decompositions of a covariance matrix, a regression-based
decomposition of the precision matrix useful in Gaussian graphical
models, a review of covariance estimation from the GLM perspective and
its evolution through linear/inverse, log and hybrid link functions.
Steinian shrinkage, regularization (banding, tapering and\break
thresholding), penalized likelihood estimation and improvement of the
sample covariance matrix for high-dimensional data are discussed in
Section~\ref{sec:shrinkest}. Some prior distributions on the parameters
of the factors of the three decompositions and their roles in the
Bayesian inference are reviewed in Section~\ref{sec:Bayesmod}.
Section~\ref{sec:what} concludes the paper.

This survey emphasizes the importance of regres\-sion-based idea and
hence the need for unconstrained reparameterization in both the GLM-
and regulariza\-tion-type approaches to covariance estimation for low-
and high-dimensional data. As such, it has a relatively narrow focus;
important topics like robustness, use of random-effects models,
nonparametric and semi-parametric methods in covariance estimation are
not discussed. It is hoped to serve as a guide or a blueprint for
further research in this active and growing area of current interest in
statistics.

\section{The GLM and Matrix Decompositions}\label{sec:overview}

In this section the importance of the GLM, the role of the three matrix
decompositions in removing the positive-definiteness constraint on a
covariance matrix, the connection between reparameterizing the
precision matrix and the Gaussian graphical models, along with linear,
log-linear and  generalized linear models for covariance matrices are
reviewed.

\subsection{Positive-Definiteness and the GLM}

A major stumbling block in covariance estimation, particularly when
using covariates, is the notorious positive-definiteness constraint.
Since a covariance matrix defined by $\Sigma = E (Y -
\mu)(Y-\mu)^\prime$, is a mean-like parameter, it is natural to exploit
the idea of GLM to develop a systematic, data-based statistical
model-fitting procedure for covariance matrices. However, unlike the
mean vector where a link function acts \textit{elementwise},
\textit{for covariance matrices elementwise transformations} are not
enough, as the positive-definiteness is a simultaneous constraint on
\textit{all} its entries. More global transformations engaging possibly
all entries of a covariance matrix are needed to remove the constraint.

Thus, the GLM approach to covariance estimation hinges on finding link
functions that induce unconstrained and statistically interpretable
reparameterization. Not surprisingly, most common and successful
modeling approaches decompose a covariance matrix into its ``variance''
and ``dependence'' components, and write regression models using
covariates for the logarithm of the ``variances.'' However, writing
such regression models for the entries of the ``dependence'' component
is still a challenging problem because these are often constrained. In
the next section examples of unconstrained\vadjust{\goodbreak} parameterizations of a
covariance matrix are given  which involve the variance-correlation,
spectral and Cholesky decompositions.

\subsection{The Matrix Decompositions}\label{subsec:md}

In this section we present the roles of the variance-correlation,
spectral and Cholesky decompositions in potentially removing the
positive-definiteness constraint on a covariance matrix, and paving the
way for using covariates to reduce its high number of parameters.

\subsubsection{The variance-correlation decomposition}

The simple decomposition $\Sigma = DRD$, where $D$ is the diagonal
matrix of standard deviations and $R = ( \rho_{ij})$ is the correlation
matrix of $Y,$ has a strong practical appeal since both factors are
easily interpreted in terms of the original variables. It allows one to
estimate $D$ and $R$ separately, which is important in situations where
one factor is more important than the other (Lin and Perlman,
\citeyear{LinPer85}; Liang and Zeger, \citeyear{LiaZeg86}; Barnard,
McCulloch and Meng \citeyear{BarMcCMen00}).

Note that while the logarithm of the diagonal entries of $D$ are
unconstrained, the correlation matrix~$R$ must be positive definite
with the additional constraint that all its diagonal entries are equal
to~$1$. Thus, it is inconvenient to work with it in the framework of
GLM and to reduce its large number of parameters. In the literature of
longitudinal data analysis (Liang and Zeger, \citeyear{LiaZeg86};
Diggle et al., \citeyear{Digetal02}; Zimmerman and
N\'{u}\~{n}ez-Ant\'{o}n, \citeyear{ZimNun10}) and other application
areas dealing with correlated data, in the interest of expediency,
parsimony and ensuring positive-definiteness structured correlation
matrices with\break a~few parameters are preferred. Fan, Huang and Li
(\citeyear{FanHuaLi07}) have studied a semiparametric model for
a~covariance structure by estimating the marginal variances via kernel
smoothing and used specific pa\-rametric models for the correlation
matrix such as the $\operatorname{ARMA}(1,1)$.

\subsubsection{Decomposition of the precision matrix: Gaussian graphical models}\label{sec:precision}

Recall that the marginal (pairwise) dependence among the entries of a
random vector is captured by the off-diagonal entries of $\Sigma$ or
the entries of the correlation matrix $R=(\rho_{ij})$.  However, the
conditional dependencies can be found in the off-diagonal entries  of
the precision matrix $\Sigma^{-1}=(\sigma^{ij}).$ More precisely, for
$Y$ a mean zero normal random vector with a positive-definite
covariance matrix, if the $ij$th component of the precision matrix is
zero, then the variables $y_i$ and $y_j$ are conditionally\vadjust{\goodbreak} independent,
given the other variables. Conditional independence structure in $Y$ is
often shown as a graphical model with the nodes corresponding to
variables and the absence of edges indicating conditional independence
(Anderson, \citeyear{And03}, Chapter~15).

In this section we give several regression interpretations of the
entries of the variance-correlation decomposition of the precision
matrix:
\[
\Sigma^{-1} = (\sigma^{ij})=\tilde D\tilde R\tilde D.
\]
Most of these are motivated by the recent surge of activities in sparse
estimation of $\Sigma^{-1}$ in the context of Gaussian graphical models
sparked by the approach in Meinshausen and  B\"{u}hlmann (\citeyear{MeiBuh06}) based
on solving $p$ separate LASSO regression problems. We show that the
entries of $(\tilde R,\tilde D)$ have direct statistical
interpretations in terms of the partial correlations, and variance of
predicting a variable given the rest. More precisely, standard
regression calculations  show that the \textit{partial correlation}
coefficient between $y_i$ and $y_j$ after removing the linear effect of
the $p-2$ remaining variables is given by $\tilde\rho_{ij}=-
\tfrac{\sigma^{ij}}{\sqrt{\sigma^{ii} \sigma^{jj}}},$ and that $\tilde
d_{ii}^2$, the \textit{partial variance} of $y_i$ after removing the
linear effect of the remaining $p-1$ variables, is given by
$\frac{1}{\sigma^{ii}}$.

For this and other regression-based techniques reviewed in this survey,
it is instructive to partition a~random vector $Y$ into two components
$(Y_1^\prime, Y_2^\prime)^\prime$ of dimensions $p_1$ and $p_2$.
Similarly, its covariance and precision matrices will be partitioned
conformally as
\[
\Sigma = \pmatrix{
\Sigma_{11} & \Sigma_{12} \cr
\Sigma_{21} & \Sigma_{22}
},
\quad
\Sigma^{-1} = \pmatrix{
\Sigma^{11} & \Sigma^{12} \cr
\Sigma^{21} & \Sigma^{22}
}.
\]

Some  useful relationships among the blocks of $\Sigma$ and
$\Sigma^{-1}$  are obtained by considering the linear least-squares
regression (prediction) of $Y_2$ based on $Y_1$. Let the $p_2 \times
p_1$ matrix $\Phi_{2|1}$ be the regression coefficients matrix and the
vector of regression residuals be denoted by $Y_{2 \cdot 1} = Y_2 -
\Phi_{2|1} Y_1$. Recall that $\Phi_{2|1}$ and the corresponding
prediction error covariance matrix can be found by requiring that the
vector of residuals $Y_{2 \cdot 1}$ be uncorrelated with $Y_1$. Thus,
\begin{equation}\label{b24}
\Phi_{2|1} = \Sigma_{21} \Sigma^{-1}_{11}=-
(\Sigma^{22})^{-1}\Sigma^{21}
\end{equation}
and
\begin{eqnarray}\label{cov24}
\operatorname{Cov} (Y_{2 \cdot 1}) &=& \Sigma_{22} - \Sigma_{21}
\Sigma^{-1}_{11} \Sigma_{12}
\nonumber \\[-8pt]\\[-8pt]
&=&\Sigma_{22 \cdot 1}=(\Sigma^{22})^{-1}.\nonumber
\end{eqnarray}

Certain special choices of $Y_2$ corresponding to $p_2 = 1,2$ are
helpful in connecting $\Phi_{2|1},\Sigma_{22 \cdot 1}$ directly to the
entries of the precision matrix $\Sigma^{-1}$, as we discuss below.

First, when $p_2 = 1$, $Y_2 = y_i$, for a fixed $i$, and~$Y_1 =\allowbreak (y_1,
\ldots, y_{i-1}, y_{i+1}, \ldots, y_p)^\prime=Y_{-(i)}$, then
$\Sigma_{22 \cdot 1}$ is a sca\-lar, called the \textit{partial variance}
of $y_i$ given the rest.  Let $\tilde y_i$ be the linear least-squares
predictor of $y_i$ based on the rest $Y_{-(i)}$, and
$\tilde\varepsilon_i=y_i-\tilde y_i$, $\tilde
d_i^2=\operatorname{Var}(\tilde\varepsilon_i)$ be its prediction error
and prediction error variance, respectively. Then,
\begin{equation}\label{reg1}
y_i=\sum_{j\neq i} \beta_{ij}y_j+\tilde\varepsilon_i,
\end{equation}
and it follows immediately from (\ref{b24}) and (\ref{cov24}) that the
regression coefficients of $y_i$ on $Y_{-(i)}$, are given by
\begin{equation}\label{regression1}
\beta_{i,j}=-\frac{\sigma^{ij}}{\sigma^{ii}},\quad j \neq i,
\end{equation}
and
\begin{equation}\label{31}
\tilde d_i^2=\operatorname{Var} (y_i | y_j, j \neq i) =
\frac{1}{\sigma^{ii}},\quad i = 1, \ldots, p.
\end{equation}
This shows that $\sigma^{ij}$, the $(i,j)$ entry of the precision
matrix, is, up to a scalar, the regression coefficient of variable $j$
in the multiple regression of variable $i$ on the rest. As such, each
$\beta_{i,j}$ is an unconstrained real number, note that
$\beta_{j,j}=0$ and $\beta_{i,j}$ is not symmetric in $(i,j)$.

Writing (\ref{regression1}) in matrix form gives another useful
factorization of the precision matrix:
\begin{equation} \label{eq:cov-}
\Sigma^{-1}= \tilde D^2( I_p- \tilde B),
\end{equation}
where $\tilde D$ is a diagonal matrix with $\tilde d_j$ as its $j$th
diagonal entry, and $\tilde B$ is a $p\times p$ matrix with zeros along
its diagonal and $\beta_{j,k}$ in the $(j,k)$th position. Now, it is
evident from (\ref{eq:cov-}) that the sparsity patterns of
$\Sigma^{-1}$ and $\tilde B$ are the same, and, hence, the former can
be inferred from the latter using the regression setup~(\ref{reg1}).
This is the key conceptual tool behind the approach of Meinshausen and
B\"{u}hlmann (\citeyear{MeiBuh06}). Note that the left-hand side of
(\ref{eq:cov-}) is a symmetric matrix while the right-hand side is not
necessarily so. Thus, one must impose the following symmetry constraint
(Rocha, Zhao and Yu \citeyear{RocZhaYu}; Friedman, Hastie and
Tibshirani, \citeyear{FriHasTib}) for $j,k=1,\ldots,p$:
\begin{equation}\label{eq:sym}
d_k^2\beta_{jk}=d_j^2\beta_{kj}.
\end{equation}

As another important example, take $p_2 = 2$, $Y_2 = (y_i, y_j)$, $i
\neq j$ and $Y_1=Y_{-(ij)}$ comprising the remaining $p-2$ variables.
Then, it follows from (\ref{cov24}) that the covariance matrix between
$y_i, y_j$, after eliminating the linear effects of the other $p-2$
components, is given by
\[
\Sigma_{22 \cdot 1} =\pmatrix{ \sigma^{ii} & \sigma^{ij} \cr
\sigma^{ij} & \sigma^{jj}}^{-1}
=
\Delta^{-1}\pmatrix{
\sigma^{jj} & - \sigma^{ij} \cr
- \sigma^{ij} & \sigma^{ii}
},
\]
where $\Delta = \sigma^{ii} \sigma^{jj} - (\sigma^{ij})^2$.
The correlation coefficient in $\Sigma_{22 \cdot 1}$ is, indeed, the
\textit{partial correlation coefficient} between $y_i$ and $y_j$:
\begin{equation}\label{pc}
\tilde\rho_{ij} = - \frac{\sigma^{ij}}{\sqrt{\sigma^{ii}}\sigma^{ij}},
\end{equation}
as announced earlier. Moreover, from (\ref{regression1}) and
(\ref{pc}) it follows that
\begin{equation}\label{regsym}
\beta_{ij}=\tilde\rho_{ij}\sqrt\frac{\sigma^{jj}}{\sigma^{ii}}.
\end{equation}
This representation which shows that $\Sigma^{-1}$ and $\tilde R$ share
the same sparsity pattern is the basis for the \citeauthor{PenZhoZhu09} (\citeyear{PenZhoZhu09}) SPACE algorithm which imposes a LASSO penalty on the
off-diagonal entries of the matrix of partial correlations $\tilde R$;
see also Friedman, Hastie and Tibshirani (\citeyear{FriHasTib}).

\subsubsection{The spectral decomposition}\label{spec}

The spectral decomposition of a covariance matrix given by
\begin{equation} \label{eq:pspec}
\Sigma = P \Lambda P^\prime=\sum_{i=1}^p \lambda_ie_ie_i^\prime,
\end{equation}
where $\Lambda$ is a diagonal matrix of eigenvalues and~$P$ the
orthogonal matrix of normalized eigenvectors with the $e_i$ as its
$i$th column, is familiar from the literature of principal component
analysis (Anderson, \citeyear{And03}; Flury, \citeyear{Flu88}). The
entries of $\Lambda$ and $P$ have interpretations as variances and
coefficients of the principal components. The matrix $P$ being
orthogonal is constrained, so that it is inconvenient to work with it
in the framework of GLM or to use covariates to reduce its high number
of parameters.

In spite of the severe constraint on the orthogonal matrix, the
spectral decomposition is the source of a new unconstrained
reparameterization due to Leonard and Hsu (\citeyear{LeoHsu92}) and
Chiu, Leonard and Tsui (\citeyear{ChiLeoTsu96}). They observed that the
logarithm of a covariance matrix $\Sigma$ defined by
\begin{equation}\label{eq:log}
\log \Sigma = P \log \Lambda P^\prime=\sum_{i=1}^p (\log \lambda_i)e_ie_i^\prime
\end{equation}
is an unconstrained symmetric matrix. However,\break a~drawback of this
transformation (link function) seems to be the lack of statistical
interpretability of the entries of $\log \Sigma$ (Brown, Le and Zidek,
\citeyear{BroLeZid94}; Liechty, Liechty and M{\"u}ller,
\citeyear{LieLieMul04}). From (\ref{eq:pspec}) and (\ref{eq:log}) it is
evident that the entries of $\Sigma$ and $\log\Sigma$ are similar
functions of the entries of $P$ and $\Lambda$, except that in~(\ref{eq:log})
$\lambda_i$ is replaced by $\log\lambda_i$. Can this
``small'' substitution be the reason for the ``big'' difference in the
statistical interpretability of the entries of log of a covariance
matrix and the matrix itself? This case is interesting as it points out
to a sort of trade-off that exists between the requirements of
unconstrained reparameterization of covariance matrices and statistical
interpretability of the new parameters.

\subsubsection{The Cholesky decompositions} \label{subsec:chol}

The standard Cholesky decomposition of a positive-definite matrix
encountered in some optimization techniques, software packages and
matrix computation (Golub and Van Loan, \citeyear{GolVan89}) is of the
form
\begin{equation} \label{eq:softw}
\Sigma = CC',
\end{equation}
where $C = (c_{ij})$ is a unique lower-triangular matrix with positive
diagonal entries. Statistical interpretation of the entries of $C$ is
difficult in its present form (Pinheiro and Bates, \citeyear{PinBat96}). However,
reducing $C$ to unit lower-triangular matrices through multiplication
by the inverse of $ D = \operatorname{diag} (c_{11}, \ldots , c_{pp}) $
makes the task of statistical interpretation of the diagonal entries of
$C$ and the ensuing unit lower-triangular matrix much easier.

For example, using basic matrix multiplication, (\ref{eq:softw}) can be
rewritten as
\begin{equation} \label{eq:Pour}
\Sigma = CD^{-1} DDD^{-1} C' = LD^2 L' ,
\end{equation}
where $L = CD^{-1}$ is obtained from $C$ by dividing the entries of its
$i$th column by $c_{ii}$. This is usually called the modified Cholesky
decomposition  of $\Sigma$; it can also be written in the forms
\begin{equation}\label{eq:Pour-}
T \Sigma T^\prime = D^2 , \quad\Sigma^{-1}=T^\prime D^{-2}T,
\end{equation}
where $T = L^{-1}$. Note that the second identity is, in fact, the
modified Cholesky decomposition of the precision matrix $\Sigma^{-1}$,
and the first identity in (\ref{eq:Pour-}) looks a lot like the
spectral decomposition, in that $\Sigma$ is diagonalized by a lower
triangular matrix. However, we show that unlike the constrained entries
of the orthogonal matrix of the spectral decomposition, the
nonredundant entries of $T = L^{-1}$ are unconstrained and
statistically  meaningful. Furthermore, the argument makes it clear
that the parameters in the factors of the Cholesky decomposition are
dependent on the \textit{order} in which the variables appear in the
random vector $Y$. Wagaman and Levina (\citeyear{WagLev09}) have
proposed an Isomap method for discovering an order among the variables
based on their correlations. This could lead to block-diagonal or
banded correlation structures which  may help to fix a reasonable order
before applying the Cholesky decomposition; see Section~\ref{sec:what}.

As in Section~\ref{sec:precision}, we use the idea of regression to
show that $T$ and $D$ can be constructed directly by regressing a
variable $y_t$ on its predecessors. In what follows, it is assumed that
$Y$ is a random vector with mean zero and a positive-definite
covariance matrix~$\Sigma$. Let $\hat{y}_t$ be the linear least-squares
predictor of $y_t$ based on its predecessors $y_{t-1}, \ldots, y_1$,
and $\varepsilon_t = y_t - \hat{y}_t$ be its prediction error with
variance $\sigma^2_t = \operatorname{Var} ( \varepsilon_t)$. Then,
there are unique scalars $\phi_{tj}$ so that
\begin{equation} \label{eq:sum3}
y_t= \sum^{t-1}_{j=1} \phi_{tj} y_j + \varepsilon_t,\quad t = 1,
\ldots, p.
\end{equation}
Next, we show how to compute the regression coefficients $\phi_{tj}$
using the covariance matrix. For a fixed~$t$, $2 \le t \le p$, set
$\phi_t=(\phi_{t1},\ldots,\phi_{t,t-1})^\prime$ and let $\Sigma_t$ be
the $(t-1)\times (t-1)$ leading principal minor of $\Sigma$ and~$\tilde
\sigma_t$ be the column vector composed of the first $t-1$ entries of
the $t$th column of $\Sigma$. Then, from (\ref{b24}) and~(\ref{cov24})
with $Y_1=(y_1,\ldots,y_{t-1})',Y_2=y_t$ it follows that
\begin{equation} \label{eq:reg}
\phi_t=\Sigma_t^{-1}\tilde\sigma_t,\quad
\sigma_t^2=\sigma_{tt}-\tilde\sigma_t^\prime
\Sigma_t^{-1}\tilde\sigma_t.
\end{equation}

Let $\varepsilon = ( \varepsilon_1, \ldots , \varepsilon_p)^\prime$ be
the vector of successive uncorrelated prediction errors with
$\operatorname{Cov} (\varepsilon ) = \operatorname{diag}
(\sigma^2_1,\allowbreak \ldots , \sigma^2_p) = D^2$. Then, ({\ref{eq:sum3})
can be rewritten in matrix form as $\varepsilon = TY$, where $T$ is the
following unit lower triangular matrix:
\begin{equation} \label{eq:ltm}
\qquad T= \pmatrix{
1 &  &  &  &  \cr
-\phi_{21} & 1 &  &  &  \cr
-\phi_{31} & -\phi_{32} & 1 &  &  \cr
\vdots &  &  & \ddots &  \cr
-\phi_{n1} & -\phi_{n2} & \cdots & -\phi_{n,n-1} & 1 \cr
}.
\end{equation}
Now, computing $\operatorname{Cov}(\varepsilon)=
\operatorname{Cov}(TY)=T\Sigma T^\prime$ gives the modified Cholesky
decomposition~(\ref{eq:Pour-}).

Since the $\phi_{ij}$'s in (\ref{eq:reg}) are simply the regression
coefficients computed from an unstructured covariance matrix, these
coefficients along with $\log \sigma^2_t$ are unconstrained
(Pourahmadi, \citeyear{Pou99,Pou00}). From (\ref{eq:sum3}) it is
evident that the regression or the orthogonalization process reduces
the task of modeling a covariance matrix to that of a sequence of $p$
varying-coefficient and varying-order regression models.\break Thus, one can
bring the familiar regression analysis machinery to handle the
unintuitive task of modeling covariance matrices (Smith and Kohn,
\citeyear{SmiKoh02}; Wu and Pourahmadi, \citeyear{WuPou03}; Huang et
al., \citeyear{Huaetal06}, Huang, Liu and Liu, \citeyear{HuaLiuLiu07};
Bickel and Levina, \citeyear{BicLev08N1}; Rothman, Levina and Zhu,
\citeyear{RotLevZhu09}). An important consequence of (\ref{eq:Pour-})
is that for any estimate $(\hat T,\hat D^2)$ of the Cholesky factors,
the estimated precision matrix $\hat\Sigma^{-1}=\hat T^\prime \hat
D^{-2}\hat T$ is guaranteed to be positive definite.

An alternative form of the Cholesky decomposition~(\ref{eq:Pour}) due
to Chen and Dunson (\citeyear{CheDun03}), also obtained from
(\ref{eq:softw}), is
\[
\Sigma=D\tilde L\tilde L^\prime D,
\]
where $\tilde L=D^{-1}C$ is obtained from $C$ by dividing the entries
of its $i$th row by $c_{ii}$. This form has proved useful for joint
variable selection for fixed and random effects in the linear
mixed-effects models, and when the focus is on modeling the correlation
matrix; see Bondell, Krishna and Ghosh (\citeyear{BonKriGho}) and
Pourahmadi (\citeyear{Pou07}).

%
%

Some early and implicit examples of the use of the Cholesky
decomposition in the literature of statistics include Bartlett's
(\citeyear{Bar33}) decomposition of a sample covariance matrix,
Wright's (\citeyear{Wri34}) path analysis, Roy's (\citeyear{Roy58})
step-down procedures and Wold's (\citeyear{Wol60}) causal chain models
which assume the existence of an \textit{a priori} order among the $p$
variables of interest. Some of the more explicit uses are in Kalman
(\citeyear{Kal60}) for filtering of state-space models and the Gaussian
graphical models (Wermuth, \citeyear{Wer80}). For other uses of
Cholesky decomposition in multivariate quality control and related
areas see Pourahmadi (\citeyear{Pou}).

\subsection{GLM for Covariance Matrices} \label{sec:glm}

\subsubsection{Linear covariance models} \label{subsec:lcm}

The origin of linear models for covariance matrices can be traced to
the work of Yule (\citeyear{Yul27}) and Gabriel (\citeyear{Gab62}) and
the implicit parameterization of a multivariate normal distribution in
terms of the entries of either $\Sigma$ or its inverse. However,
Dempster (\citeyear{Dem72}) was the first to recognize the entries of
$\Sigma^{-1}=( \sigma^{ij})$ as the canonical parameters of the
exponential family of normal distributions. He proposed to select or
estimate a covariance matrix efficiently and sparsely by identifying
zeros in its inverse, and referred to the procedure as
\textit{covariance selection} models. It fits the framework of linear
covariance models defined next.

Motivated by the simple and linear structure of covariance matrices of
some time series and variance component models, Anderson
(\citeyear{And73}) introduced the class of \textit{linear covariance
models} (LCM):
\begin{equation}\label{lcm}
\Sigma^{\pm 1} = \alpha_1 U_1 + \cdots + \alpha_q U_q,
\end{equation}
where $U_i$'s are some known symmetric basis matrices and $\alpha_i$'s
are unknown parameters; they must be restricted so that the matrix is
positive definite. It is usually assumed that there is at least a set
of coefficients where $\Sigma^{\pm 1}$ is positive definite. The model
(\ref{lcm}) is rather general, indeed, for $q=p^2$ any covariance
matrix admits the representation
\begin{equation} \label{eq:as}
\Sigma = (\sigma_{ij}) = \sum^p_{i=1} \sum^p_{j=1} \sigma_{ij}
U_{ij}, 
\end{equation}
where $U_{ij}$ is a $p \times p$ matrix with one on the $(i,j)$th
position and zero elsewhere.

Replacing $\Sigma$ by $S$ in the left-hand side of (\ref{lcm}), it can
be viewed as a collection of $\frac {p(p+1)}{2}$ linear regression
models. The same regression models viewpoint holds with the precision
matrix on the left-hand side. The class of linear covariance models is
omnipresent when dealing with covariance matrices. It includes
virtually any estimation method that acts elementwise on a covariance
matrix such as tapering, banding, thresholding, covariance selections
models, penalized likelihood with LASSO penalty on the entries of the
precision matrix, etc.; see (\ref{eq:as}).

A major drawback of (\ref{lcm}) and (\ref{eq:as}) is the constraint on
the coefficients which could make the estimation and other statistical
problems difficult (Anderson, \citeyear{And73}). Szatrowski
(\citeyear{Sza80}) gives necessary and sufficient conditions for the
existence of explicit maximum likelihood estimates, and the convergence
of the iterative procedure in one iteration from any positive-definite
starting point.

A good review of the MLE procedures for the model~(\ref{lcm}) and their
applications to the problem of testing homogeneity of the covariance
matrices of several dependent multivariate normals is presented in
Jiang, Sarkar and Hsuan (\citeyear{JiaSarHsu99}). They derive
a~likelihood ratio test, and show how to compute the MLE of $\Sigma$, in
both the restricted (null) and unrestricted (alternative) parameter
spaces using SAS PROC MIXED software. They also provide the code and
the implementation is explained using several examples.

The notion of covariance regression introduced by Hoff and Niu
(\citeyear{HofNiu}) is also in the spirit of (\ref{lcm}), but unlike
the LCM the covariance matrix is quadratic in the covariates, and
positive-definiteness is guaranteed through the special construction.

\subsubsection{Log-linear covariance models} \label{subsec:loglin}

A plausible\break way~to remove the constraint on $\alpha_i$'s in (\ref{lcm})
is to work with the logarithm of a covariance matrix. The key fact
needed here is that for a general covariance matrix with the spectral
decomposition $\Sigma = P \Lambda P^\prime$, its \textit{matricial
logarithm} defined by $\log \Sigma = P \log \Lambda P^\prime$ is
a~symmetric matrix with unconstrained entries taking values in $(-
\infty, \infty)$.

This  idea has been pursued by Leonard and Hsu (\citeyear{LeoHsu92})
and Chiu, Leonard and Tsui (\citeyear{ChiLeoTsu96}) who introduced the
\textit{log-linear covariance} models for $\Sigma$ as
\begin{equation}\label{llm}
\log \Sigma = \alpha_1 U_1 + \cdots + \alpha_q U_q,
\end{equation}
where $U_i$'s are known matrices as before and the~$\alpha_i$'s are now
unconstrained. However, since $\log\!\Sigma$ is a~high\-ly nonlinear
operation on $\Sigma$, the $\alpha_i$'s lack statistical interpretation
(Brown, Le and Zidek, \citeyear{BroLeZid94}; Liechty, Liechty and
M{\"u}ller, \citeyear{LieLieMul04}). Fortunately, for~$\Sigma$ diagonal
since $\log \Sigma = \operatorname{diag} ( \log \sigma_{11}, \ldots,
\log \sigma_{pp})$ is also diagonal, it can be seen that (\ref{llm})
amounts to log-linear models for heterogeneous variances which have
a~long history in econometrics and other areas; see Carroll and Ruppert
(\citeyear{CarRup88}) and references therein.

Maximum likelihood estimation procedures for the parameters in
(\ref{llm}) and their asymptotic properties are studied in Chiu,
Leonard and Tsui (\citeyear{ChiLeoTsu96}) along with the analysis of
two real data sets. Given the flexibility of the log-linear models, one
would expect them to be used widely in practice, however, this does not
seem to be the case. An interesting application to spatial
autoregressive (SAR) models and some of its computational advantages
are discussed in LeSage and Pace (\citeyear{LeSPac07}).

\vspace*{3pt}\subsubsection{GLM via the Cholesky decomposition} \label{subsec:GLM}

In this section the constraint and lack of interpretation\break
of~$\alpha_i$'s in (\ref{lcm}) and (\ref{llm}) are resolved simultaneously
by relying on the Cholesky decomposition of a covariance matrix
described in Section~\ref{subsec:chol}. A bona fide GLM for the
precision matrix in terms of covariates is introduced and its maximum
likelihood estimation (MLE) is discussed.  An important consequence of
the approach based on the modified Cho\-lesky decomposition is that for
any estimate  of the Cholesky factors, the estimated precision
matrix\break $\hat\Sigma^{-1}=\hat T^\prime \hat D^{-2}\hat T$ is guaranteed to be
positive definite.

Recall that for an unstructured covariance matrix $\Sigma$, the
nonredundant entries of its components $(T, \log D^2)$ in
(\ref{eq:Pour-}) are unconstrained. Thus, following the GLM's
tradition, one may write parameteric models for them using covariates
(Pourahmadi, \citeyear{Pou99}; Pan and MacKenzie, \citeyear{PanMac03};
Zimmerman and N\'{u}\~{n}ez-Ant\'{o}n, \citeyear{ZimNun10}). We
consider the following parametric models for $\phi_{tj}$ and $\log
\sigma^2_t$, for $t = 1, \ldots, p$; $j = 1, \ldots, t-1$,
\begin{equation}\label{9}
\log \sigma^2_t = z^\prime_t \lambda,\quad \phi_{tj} = z^\prime_{tj}
\gamma.
\end{equation}
Here, $z_t, z_{tj}$ are $q \times 1$ and $d \times 1$ vectors of known
covariates,  $\lambda = (\lambda_1, \ldots, \lambda_q)^\prime$ and
$\gamma = (\gamma_1, \ldots, \gamma_d)^\prime$ are parameters related
to the innovation variances and dependence in $Y$, respectively
(Pourahmadi, \citeyear{Pou99}). The most common covariates used in the
analysis of several real longitudinal data sets (Pourahmadi,
\citeyear{Pou99}; Pourahmadi and Daniels, \citeyear{PouDan02}; Pan and
Mac\-Kenzie, \citeyear{PanMac03}; Lin and
Wang, \citeyear{LinWan09}; Leng, Zhang and Pan, \citeyear{LenZhaPan10})
are in terms of powers of times and lags
\begin{eqnarray*}
z_t&=&(1,t,t^2,\ldots,t^{d-1})^\prime,
\\
z_{tj}&=&\bigl(1,t-j,(t-j)^2,\ldots,(t-j)^{p-1}\bigr)^\prime.
\end{eqnarray*}

A truly remarkable feature of (\ref{9}) is its flexibility in reducing
the potentially high-dimensional and constrained parameters of $\Sigma$
or the precision matrix to $q + d$ unconstrained parameters $\lambda$
and $\gamma$.  Furthermore, one can rely on graphical tools such as the
regressogram or AIC to identify models such as~(\ref{9}) for the data;
for more details see Pourahmadi (\citeyear{Pou99,Pou01}) and Pan and
MacKenzie (\citeyear{PanMac03}). Liang
and Zeger (\citeyear{LiaZeg86}) employ
such parametrized models for covariance matrices in the context of the
popular generalized estimating equations for longitudinal data.

Computing the MLE of the parameters is relatively simple due to the
special form of the loglikelihood function for a sample
$Y_1,\ldots,Y_n$ from a normal population with mean zero and the common
covariance $\Sigma$ parameterized as in (\ref{9}). Note that, except
for a constant, we have
\begin{eqnarray} \label{eq:loglik}
-2 l(\lambda,\gamma) &= &\sum_{i=1}^n (\log |\Sigma|
+ Y_i' \Sigma^{-1} Y_i)  \nonumber  \\
&=& n \log |D^2|+ n\operatorname{tr} \Sigma^{-1} S \nonumber \\
& =& n \log |D^2| +n \operatorname{tr} D^{-2}TST^\prime   \\
&=& n \log |D^2|\nonumber \\
&&{} +n \operatorname{tr} D^{-2}(I_p-B)S(I_p-B)^\prime,\nonumber
\end{eqnarray}
where $S=\frac {1} {n} \sum_{i=1}^n Y_iY_i^\prime$, $B=I_p-T$ and the
last three equalities are obtained by replacing for $\Sigma^{-1}$ from
(\ref{eq:Pour-}) and some basic matrix operations involving trace of a
matrix. Since (\ref{eq:loglik}) is quadratic in $B$, for a given $D^2$
the MLE of $B$ or $\phi_{tj}$'s has a closed form, the same is true of
the MLE of $D^2$ for a given~$B$ (Pourahmadi, \citeyear{Pou00}; Huang et al.,
\citeyear{Huaetal06}; Huang, Liu and Liu, \citeyear{HuaLiuLiu07}). This
simplicity in computing the MLE of the saturated (unstructured) model
for $(T,\allowbreak D)$ is important when comparing the computational aspects of
Cholesky-based estimation of the precision matrix with the Rocha, Zhao
and Yu (\citeyear{RocZhaYu}) SPLICE algorithm; see
Section~\ref{sss:PL}.

An algorithm for computing the MLE of the parameters $(\gamma,\lambda)$
using the iterative Newton--Raphson algorithm with Fisher scoring is
given in Pourahmadi (\citeyear{Pou00}) along with the asymptotic
properties of the estimators. An unexpected finding is the asymptotic
orthogonality of the MLE of the parameters $\lambda$ and $\gamma$, in
the sense that their Fisher information matrix is block-diagonal; see
Pourahmadi (\citeyear{Pou07}) and
references therein. When the assumption of normality is questionable
like when the data exhibit thick tails, then a multivariate
$t$-distribution might be a~reasonable alternative; see Lin and Wang
(\citeyear{LinWan09}) and Lin (\citeyear{Lin}).

\section{Regularization of the Sample Covariance Matrix} \label{sec:shrinkest}

This section is devoted to high-dimensional data where the sample
covariance matrix is known to be a poor estimator, and not even
invertible when $p\gg n$. We review some alternative and improved
estimators obtained by regularizing the sample covariance matrix in
various ways. After presenting a~few loss functions in
Section~\ref{subsec:loss}, we review in
Sections~\ref{subsec:spde}~and~\ref{subsec:Led} shrinkage estimators obtained by
minimizing certain risk functions. An early and inspiring example is
the Stein's family of shrinkage estimators that shrinks the eigenvalues
of the sample covariance matrix toward a central value. Penalized
normal likelihood estimators with a LASSO penalty on the precision
matrix are reviewed in Section~\ref{sss:PL} with a focus on the
graphical LASSO algorithm. Regularization methods which act elementwise
on the sample covariance matrix such as tapering, banding and
thresholding are discussed in Section~3.5. Some conditions for
consistency of such estimators are also reviewed.

\subsection{Some Loss and Risk Functions} \label{subsec:loss}

Regularized estimators are usually obtained by minimizing suitable
norms, risks or objective functions. For covariance matrix estimation
the Frobenius and operator (spectral) norms are quite natural and have
proved useful in establishing theoretical properties of covariance
estimators. For example,
 consistency in operator norm
guarantees the consistency of the eigenstructure used in principal
component analysis (Johnstone and Lu, \citeyear{JohLu09}) and other related methods
in multivariate statistics; see Section~\ref{schur}.

The two commonly used loss functions when $n > p$ are
\begin{eqnarray*}
L_1 ( \hat{\Sigma}, \Sigma ) &=& \operatorname{tr} ( \hat{\Sigma} \Sigma^{-1} ) - \log
|
\hat{\Sigma} \Sigma^{-1} | - p,
\\
L_2 ( \hat{\Sigma}, \Sigma) &=& \operatorname{tr} ( \hat{\Sigma} \Sigma^{-1} - I)^2,
\end{eqnarray*}
where $\hat{\Sigma} = \hat{\Sigma} (S)$ is an arbitrary estimator. The
corresponding risk functions are
\[
R_i ( \hat{\Sigma}, \Sigma ) = E_\Sigma L_i ( \hat{\Sigma}, \Sigma ),\quad i
= 1,2 .
\]
An estimator $\hat{\Sigma}$ is considered better than $S$ if its risk
function is smaller than that of $S$. The loss function~$L_1$ was
advocated by Stein (\citeyear{Ste56}) and is usually called the entropy
loss or the Kullback--Liebler divergence of two multivariate normal
densities corresponding to the two covariance matrices. The second,
called a quadratic loss function, is essentially the Euclidean or the
Frobenius norm of its matrix argument which involves squaring the
difference between aspects of the estimator and the target.
Consequently, it penalizes overestimates more than underestimates, and
``smaller''  estimates are more favored under $L_2$ than under $L_1$.
For example, among all scalar multiples $aS$ of the sample covariance
matrix, it is known (Haff, \citeyear{Haf80}) that $S$ is optimal under~$L_1$,
while the smaller estimator $\frac {n S}{n+p+1}$ is optimal
under $L_2$.

Following the lead of Muirhead and Leung (\citeyear{LeuMui87}), Ledoit
and Wolf (\citeyear{LedWol04}) have used a slight modification of the
Frobenius norm as the loss function
\[
L_3 ( \hat{\Sigma}, \Sigma ) =p^{-1} \| \hat{\Sigma} - \Sigma \|^2 =
p^{-1}\operatorname{tr} ( \hat{\Sigma} - \Sigma )^2.
\]
Note that though dividing by the dimension $p$ is not standard, it has
the advantage that norm of the identity matrix is one, regardless of
the size of $p$. Also, the loss $L_3$ does not involve matrix inversion
which is ideal with regard to computational cost for the ``small $n$,
large $p$'' case. The heuristics behind this loss function are the same
as those for~$L_2$. However, it has an additional and attractive
feature that the optimal covariance estimator under $L_3$ turns out to
be the penalized normal likelihood estimator with $\operatorname{tr}
\Sigma^{-1}$ as the penaly (Warton, \citeyear{War08}; Yuan and Huang,
\citeyear{YuaHua09}). Since the penalty function becomes large when
$\Sigma$ gets closer to singularity, such a penalty forces the
covariance estimators to be nonsingular and better conditioned.



\subsection{Shrinking the Spectrum and the Correlation Matrix} \label{subsec:spde}

In this section we present one of the earliest improvements of $S$
obtained by shrinking only its eigenvalues. Having observed that the
sample covariance matrix systematically distorts the
eigenstructure\break
of~$\Sigma$, particularly when $\frac{p}{n}$ is large, Stein
(\citeyear{Ste56,Ste75}) initiated the task of improving it. He
considered orthogonally invariant estimators of the form
\[
\hat{\Sigma} = \hat{\Sigma} (S) = P \Phi ( \lambda ) P^\prime ,
\]
where $\lambda = ( \lambda_1 , \ldots , \lambda_p)^\prime, \lambda_1 >
\cdots > \lambda_p > 0$, are the ordered eigenvalues of $S$, and $P$ is
the orthogonal matrix whose $j$th column is the normalized eigenvector
of $S$ corresponding to $\lambda_j$, and
$\Phi ( \lambda )\,{=}\,\operatorname{diag} (\!\varphi_1 , \ldots , \varphi_p\!)$ is a diagonal
matrix where $\varphi_j = \varphi_j ( \lambda )$ estimates the $j$th
largest eigenvalue of $\Sigma$. For example, the choice of $\varphi_j =
\lambda_j$ corresponds to the usual unbiased estimator $S$, where it is
known that $\lambda_1$ and $\lambda_p$ have upward and downward biases,
respectively. Stein's method chooses $\Phi ( \lambda )$ so as to
counteract the biases of the eigenvalues of $S$ by shrinking them
toward some central values. For the $L_1$ risk, his modified estimators
of the eigenvalues of $\Sigma$ are $ \varphi_j =
\frac{n\lambda_j}{\alpha_j}, $ where
\[
\alpha_j = \alpha_j ( \lambda ) = n-p+1+2  \lambda_j \sum_{i \neq j}
\frac{1}{\lambda_j - \lambda_i}.
\]
Note that the $\varphi_j$'s will differ the most from $\lambda_j$ when
some or all of the $\lambda_j$'s are nearly equal and $\frac {n}{p}$ is
not small. Since some of the $\varphi_j$'s could be negative and may
not even satisfy the order restriction, Stein has suggested an
isotonizing procedure to obtain modified estimators satisfying the
above constraints; for more details on this procedure see Lin and
Perlman (\citeyear{LinPer85}).

Lin and Perlman (\citeyear{LinPer85}) have applied the James--Stein
shrinkage estimators (James and Stein, \citeyear{JamSte61}) to the
sample correlation in order to improve it for large~$p$. They shrink
the Fisher $z$-transform of the individual correlation coefficients
(and the logarithm of the variances) toward a common target value.

\vspace*{3pt}\subsection{Ledoit--Wolf Shrinkage Estimator}\vspace*{3pt}
\label{subsec:Led}

To ensure nonsingularity of the estimated covariance matrix in the
``$n$ small, $p$ large'' case, Ledoit and Wolf (\citeyear{LedWol04})
present a shrinkage estimator that is asymptotically the optimal convex
linear combination of the sample covariance matrix and the identity
matrix with respect to $L_3$.

One can motivate such an estimator by recalling that the sample
covariance matrix $S$ is unbiased for~$\Sigma$, but unstable with
considerable risk when $p\gg n$. By contrast, a structured covariance
matrix estimator like the identity matrix has very little estimation
error, but can be severely biased when the structure is misspecified. A
natural compromise between these two extremes is a linear combination
of them, giving a simple shrinkage or ridge candidate of the form
\[
\hat{\Sigma} = \alpha_1 I +  \alpha_2 S.
\]
Now, one may choose $\alpha_1$ and $\alpha_2$ to optimize certain
criterion (Ledoit and Wolf, \citeyear{LedWol04}).

Using the Frobenius norm or minimizing the risk corresponding to the
loss function $L_3$, Ledoit and Wolf (\citeyear{LedWol04}) showed that
the optimal choices of $\alpha_1$ and $\alpha_2$ depend only on the
following four-dimensio\-nal aspects of the true (but unknown) covariance
matrix~$\Sigma$:
\begin{eqnarray*}
\mu&=&\operatorname{tr} (\Sigma)/p , \quad \alpha^2= \|\Sigma-\mu
I\|^2,
\\
\beta^2&=&E\|S-\Sigma\|^2 ,\quad \delta^2=E\|S-\mu I\|^2.
\end{eqnarray*}
Consistent estimators of these low-dimensional parameters are provided
by Ledoit and Wolf (\citeyear{LedWol04}), so that substitution in
$\hat{\Sigma}$ results in a positive-definite estimator of $\Sigma$.
Through extensive simulation studies they establish the superiority of
this estimator to the sample covariance matrix and the empirical Bayes
estimator (Haff, \citeyear{Haf80}), among others.

Warton (\citeyear{War08}) taking $\alpha_2=1$ showed that such ridge
estimators can be obtained using the penalized normal likelihood where
the penalty term is proportional to $\operatorname{tr}\Sigma^{-1}$.
Evidently, such a penalty ensures that the estimator is a nonsingular
matrix. He suggests using the cross-validation of the likelihood
function for estimation of the ridge and the penalty parameters, and
extends the approach to the ridge estimation of the correlation matrix.
His method of estimation leads to the definition of suitable test
statistics for the parameters in multivariate linear regression in
high-dimensional situations. The power properties of the test statistic
are studied and compared with the principal components and generalized
inverse test statistics used in dealing with high dimensionality.

\subsection{The Penalized Likelihood Approach} \label{sss:PL}

In this section we review various regularization methods based on
penalizing the normal likelihood. These methods differ mostly on the
LASSO\vadjust{\goodbreak} penalty imposed on certain segments of the precision matrix. For
example, Huang et al. (\citeyear{Huaetal06}), Banerjee, El~Ghaoui and
d'Aspremont (\citeyear{BanElGdAs08}), Friedman, Has\-tie and Tibshirani
(\citeyear{FriHasTib08}), Rothman et al. (\citeyear{Rotetal08}) and
Warton (\citeyear{War08}), respectively, impose penalty on the Cholesky
factor, all the entries, off-diagonal entries and the diagonal entries
of the precision matrix. These can be viewed as methods for solving
Dempster's (\citeyear{Dem72}) covariance selection problem of inducing
sparsity in the precision matrix. However, War\-ton's (\citeyear{War08})
penalty leads to the Ledoit--Wolf estimator where neither $\Sigma$ nor
its inverse is sparse.

Motivated by the success of the LASSO estimators in the context of
linear regression with a large number of covariates (Tibshirani,
\citeyear{Tib96}), and in view of~(\ref{lcm})~and~(\ref {eq:as}), it is
plausible to induce sparsity in the precision matrix estimate by adding
to the normal loglikelihood (\ref{eq:loglik}) a penalty on the entries
of the precision matrix $\Sigma^{-1}$ or its Cholesky factor (Huang et
al., \citeyear{Huaetal06})
\begin{equation} \label{eq:penaly}
-2l+ \sum_{i<j}p_{\lambda_{ij}}(\sigma^{ij}),
\end{equation}
where $\sigma^{ij}$ is the $(i,j)$th entry of the precision matrix and
$\lambda_{ij}$ is the corresponding tuning parameter. Note that the
LASSO penalty corresponds to $ p_{\lambda}(|x|) = \lambda |x|$. Such an
approach will inherit many desirable computational and statistical
properties of LASSO and its many improved variants (Efron et~al.,
\citeyear{Efretal04}; Rocha, Zhao and Yu, \citeyear{RocZhaYu}; Fan and
Lv, \citeyear{FanLv10}, Section~3.5).

Some early attempts at inducing sparsity in the precision matrix are
Bilmes (\citeyear{Bil00}), Smith and Kohn (\citeyear{SmiKoh02}), Wu and
Pourahmadi (\citeyear{WuPou03}) and Levina,\break Rothman and Zhu
(\citeyear{LevRotZhu08})
who, for a fixed order of the variables in $Y,$ use a parametrization
of the precision matrix in terms of the modified Cholesky
decomposition~(\ref{eq:Pour-}). Covariance selection priors and AIC
were used to promote sparsity in $T$. Huang et~al.
(\citeyear{Huaetal06}) proposed a covariance selection estimator  by
adding to the normal loglikelihood the LASSO penalty on the
off-diagonal entries of $T, $ and cross-validation was used to select a
common regularization parameter; see also Huang, Liu and Liu
(\citeyear{HuaLiuLiu07}) and Levina, Rothman and Zhu (\citeyear{LevRotZhu08}) for some
improvements. Bickel and Levina (\citeyear{BicLev08N1}) provide
conditions ensuring consistency in the operator norm for the precision
matrix estimates based on banded Cholesky factors.

Chang and Tsay (\citeyear{ChaTsa10}) extend the Huang et al.
(\citeyear{Huaetal06}) setup using an equi-angular penalty which
imposes different penalty on each row of $T$ and the penalties are
inversely proportional to the prediction variance $\sigma_t^2$ of the
$t$th regression. Extensive simulations were used to compare the
performance of their method with others, including the sample
covariance matrix, banding (Bickel and Levina, \citeyear{BicLev08N1})
and the $L_1$-penalized normal loglikelihood (Huang et al.,
\citeyear{Huaetal06}). Contrary to the banding method, the method of
Huang et al. and the equi-angular method worked reasonably well for six
covariance matrices, with the equi-angular method outperforming the
others. Since the modified Cholesky decomposition is not
permutation-invariant, they also use a random permutation of the
variables before estimation to study the sensitivity to permutation of
each method. They conclude that permuting the variables introduces some
difficulties for each estimation method, except the sample covariance
matrix, but the equi-angular method remains the best, with the banding
method having the worst sensitivity to permutation. They also compare
these methods by applying them to a portfolio selection problem with
$p=80$ series of actual daily stock returns.

Two disadvantages of imposing sparsity on the~fac\-tor $T$ are that its
sparsity does not necessarily \mbox{imply} sparsity of the precision matrix,
and  the sparsity structure in $T$ could be sensitive to the order of
the random variables within $Y$. Some alternative methods which tackle
these issues penalize the precision matrix directly. For example,
Banerjee, El~Ghaoui and d'Aspremont (\citeyear{BanElGdAs08}), Yuan and
Lin (\citeyear{YuaLin07}) and Friedman, Hastie and Tibshirani
(\citeyear{FriHasTib08}) consider an estimate defined by the normal
loglikelihood penali\-zed by the $L_1$-norm of the entries
of~$\Sigma^{-1}$. These~me\-thods produce sparse, permutation-invariant
estimators of the precision matrix, though some are computationally
expensive. Yuan and Lin (\citeyear{YuaLin07}) used the max-det
algorithm to compute the estimator while imposing the
positive-definiteness constraint; this seems to have limited their
numerical results to $p \le 10$ (Rothman et al. \citeyear{Rotetal08},
page~496).

To date, the fastest available algorithm is the gra\-phical lasso
(glasso), proposed by Friedman, Hastie and Tibshirani
(\citeyear{FriHasTib08}). It relies on the equivalence of the Banerjee,
El~Ghaoui and d'Aspremont (\citeyear{BanElGdAs08}) blockwise interior
point procedure and recursively solving and updating a series of LASSO
regression problems using the coordinate descent algorithm for LASSO.
Fortunately, the sparse covariance estimator from the graphical LASSO
is guaranteed to be~po\-sitive definite. This important property follows
from a result due to Banerjee, El~Ghaoui and d'Aspremont
(\citeyear{BanElGdAs08}) showing that if the iterative procedure is
initialized with a positive-definite matrix, then the subsequent
iterates remain positive definite.

The sparse pseudo-likelihood inverse covariance estimation (SPLICE)
algorithm of Rocha, Zhao and Yu (\citeyear{RocZhaYu}) and the SPACE
(Sparse PArtial Correlation Estimation) algorithm of Peng et al.
(\citeyear{PenZhoZhu09}) also impose sparsity constraints directly on
the precision matrix, but with slightly different regression-based
reparameterizations of $\Sigma^{-1}$; see (\ref{eq:cov-}) and
(\ref{pc}). They are designed to improve several shortcomings of the
approach of Meinshausen and B\"{u}hlmann (\citeyear{MeiBuh06}), including
its lack of symmetry for neighborhood selection in Gaussian graphical
models. While Meinshausen and B\"{u}hlmann (\citeyear{MeiBuh06}) use $p$
separate linear regressions to estimate the neighborhood of one node at
a time, Rocha et al. and Peng et al. propose merging all p linear
regressions into a single least squares problem where the observations
associated to each regression are weighted according to their
conditional variances.

To appreciate the need for using approximate or pseudo-likelihood, it
is instructive to note that unlike the sequence  of prediction errors
in (\ref {eq:sum3}), the $\tilde\varepsilon_j$'s from
Section~\ref{sec:precision} are correlated so that $\tilde D^2$ is not
really the covariance matrix of the vector of regression errors
$\tilde\varepsilon=(\tilde\varepsilon_1,\ldots,\tilde\varepsilon_p)^\prime$.
The use of its true and full covariance matrix in the normal
loglikelihood would increase the computational cost at the estimation
stage. This problem is circumvented in Rocha, Zhao and Yu
(\citeyear{RocZhaYu}) and Friedman, Hastie and Tibshirani
(\citeyear{FriHasTib}) by using a pseudo-likelihood function which in
the normal case amounts to pretending that the $\operatorname{Cov}
(\tilde\varepsilon)$ is $\tilde D^2$. To this pseudo-loglikelihood
function, they add the symmetry constraints (\ref{eq:sym}) and
a~weighted LASSO penalty on the off-diagonal entries to promote sparsity.
A drawback of the SPLICE and SPACE algorithms is that they do not
enforce the positive-definiteness constraint, hence, the resulting
covariance estimators are not guaranteed to be positive definite.

The \textit{sparsistency} and rates of convergence for spar\-se
covariance and precision matrix estimation using the penalized
likelihood with nonconvex penalty functions have been studied in Lam
and Fan (\citeyear{LamFan09}). Sparsistency refers to the property that
all zero entries  are actually estimated as zero with probability
tending to one. In a given situation, sparsity might be present in the
covariance matrix, its inverse or Cholesky factor. They develop a
unified framework to study these three sparsity problems with a general
penalty function and show that the rates of convergence for these
problems under the Frobenius norm are of the order $(\frac {s\log
p}{n})^{1/2}$, where $s=s_n$ is the number of nonzero elements, $p=p_n$
is the size of the covariance matrix and $n$ is the sample size. This
reveals that the contribution of high-dimensionality is merely of a
logarithmic factor.

\vspace*{3pt}\subsection{Elementwise Shrinkage}\vspace*{3pt}
\label{schur}

In this section we review a few alternative estimators like
\textit{banding}, \textit{tapering} and \textit{thresholding} which are
based on the elementwise shrinkage of the sample covariance matrix.
These covariance estimators require a minimal amount of computation,
except in the cross-validation for selecting the tuning parameter which
is computationally comparable to that for the penalized likelihood
method. However, due to their emphasis on elementwise transformations,
such estimators are not guaranteed  to be positive definite.

\subsubsection{Banding and tapering the sample covariance matrix} \label{sss:BSCM}

Many entries of the sample covariance matrix $S=(s_{ij})$ could be
small or unstable in the\break ``$n$~small, $p$~large'' case. The most
extreme case of this occurs in time series analysis where one has to
work with only a single (long) realization $(n = 1)$. The requirement
of stationarity reduces the number of distinct entries of the $p \times
p$ covariance matrix~$\Sigma$ from $p(p+1)/2$ to just $p$, which is
still large. The moving average (MA) and autoregressive (AR) models
which further reduce the number of parameters are the prototypes of
banding a covariance/precision matrix (Bickel and Levina,
\citeyear{BicLev04}; Wu and Pourahmadi, \citeyear{WuPou09}; McMurry and
Politis, \citeyear{McMPol10}).

Given the sample covariance matrix $S = (s_{ij})$ and any integer $k$, $0
\leq k < p$, its $k$-banded (Bickel and Levina, \citeyear{BicLev08N1})
version defined by
\[
B_k (S) = [s_{ij} {\mathbf{1}}(|i - j | \leq k)]
\]
can serve as an estimator for $\Sigma$. This kind of regularization is
ideal when the indices have been arranged so that
\[
|i - j| > k \quad\Longrightarrow\quad \sigma_{ij} = 0.
\]
This occurs, for example, if $y_1, y_2, \ldots ,$ form a finite
inhomogenous moving average process
\[
y_t = \sum^k_{j=1} \theta_{t,t-j} \varepsilon_j,
\]
and $\varepsilon_j$'s are i.i.d. with mean $0$ and finite\vadjust{\goodbreak} variances.

Banding is a special case  of tapering which repla\-ces $S$ by $S * R$,
where ($*$) denotes the Schur (coordi\-nate-wise) matrix multiplication
and $R = (r_{ij})$ is a~positive-definite symmetric matrix (Furrer
and\break
Bengtsson, \citeyear{FurBen07}). It is known that the Schur product of
two positive-definite matrices is also positive definite. Banding
corresponds to using $R=r_{ij} = ({\mathbf{1}}(|i-j| \leq k)$, which is
not a positive-definite matrix. The idea of banding has also been used
on the lower triangular matrix of the Cholesky decomposition of
$\Sigma^{-1}$ by Wu and Pourahmadi (\citeyear{WuPou03}), Huang et al.
(\citeyear{Huaetal06}) and Bickel and Levina (\citeyear{BicLev08N1}).
While Furrer and Bengtsson (\citeyear{FurBen07}) have used tapering as
a regularization technique for the ensemble Kalman filter, Kaufman,
Schervish and Nychka (\citeyear{KauSchNyc08}) use it for purely
computational purposes in the likelihood-based estimation of the
parameters of a structured covariance function for large spatial data
sets.

Asymptotic analysis of banding is possible when~$n$, $p$ and $k$ are
large. Bickel and Levina [(\citeyear{BicLev08N1}), Theorems 1 and 2] have
shown that, for normal data, the banded estimator is consistent in the
operator norm (spectral norm), uniformly over a class of approximately
``bandable'' matrices, as long as \mbox{$\frac{\log p}{n} \rightarrow 0$}.
They obtain explicit rate of convergence which depends on how fast $k
\rightarrow \infty$; see also Cai, Zhang and Zhou
(\citeyear{CaiZhaZho10}). The consistency in operator norm guarantees
the consistency of principal component analysis (Johnstone and Lu,
\citeyear{JohLu09}) and other related methods in multivariate
statistics when $n$ is small and $p$ is large. Cai, Zhang and Zhou
(\citeyear{CaiZhaZho10}) propose a tapering procedure for the
covariance matrix estimation and derive the optimal rate of convergence
for estimation under the operator norm. They also carry out a
simulation study to compare the finite sample performance of their
proposed estimator with that of the banding estimator introduced in
Bickel and Levina (\citeyear{BicLev08N1}). The simulation shows that
their proposed estimator has good numerical performance, and nearly
uniformly outperforms the banding estimator.

\subsubsection{Thresholding the sample covariance matrix} \label{sss:TSCM}

When both $n$ and $ p$ are large, it is plausible that many elements of
the population covariance matrix are equal to $0$, and, hence, $\Sigma$
is sparse. How does one develop an estimator other than $S$ to cope
with this situation? The concept of thresholding originally developed
in nonparametric function estimation has been used in the estimation of
large covariance matrices by Bickel and Levina (\citeyear{BicLev08N2}),
El Karoui (\citeyear{ElK08N1,ElK08N2}) and Rothman, Levina and
Zhu\vadjust{\goodbreak}
(\citeyear{RotLevZhu09}).

For a sample covariance matrix $S = (s_{ij})$ the thres\-holding operator
$T_s$ for $s \geq 0$ is defined by
\[
T_s (S) = [ s_{ij} {\mathbf{1}} (|s_{ij} | \geq s) ].
\]
Thus, thresholding $S$ at $s$ amounts to replacing by zero all entries
with absolute value less than $s$. Its biggest advantage is its
simplicity, as it carries no~ma\-jor computational burden compared to its
competitors like the penalized likelihood with the LASSO penalty (Huang
et al., \citeyear{Huaetal06}; Rothman et al., \citeyear{Rotetal08};
Friedman, Hastie and Tibshirani, \citeyear{FriHasTib08}). A potential
disadvantage is the loss of positive-definiteness as in banding.
However, just as in banding, Bickel and Levina (\citeyear{BicLev08N2})
have established the consistency of the threshold estimator in the
operator norm, uniformly over the class of matrices that satisfy a
notion of sparsity, provided that $\frac{\log p}{n} \rightarrow 0$. An
immediate consequence of the consistency result is that a threshold
estimator will be positive definite with probability tending to one for
large samples and dimensions.

\section{Bayesian Modeling of Covariances} \label{sec:Bayesmod}

Heuristically, there is an implicit equivalence between  regularization
and Bayesian estimation in sta\-tistics. This can be seen by suitable
exponentiation of the penalty term in (\ref{eq:penaly}) and viewing it
as a prior on the parameter space, or conversely by viewing a prior as
a means of imposing constraints on the parameters.

Traditionally, in Bayesian approaches to inference for $\Sigma$ the
Jefferys' improper prior and the conjugate inverse Wishart (IW) priors
are used. For some reviews of the earlier work in this direction, see
Lin and Perlman (\citeyear{LinPer85}) and Brown, Le and Zidek
(\citeyear{BroLeZid94}). However, the success of Bayesian computation
and Markov Chain Monte Carlo (MCMC) in the late 1980s did open up the
possibility of using more flexible and elaborate nonconjugate priors
for covariance matrices; see Leonard and Hsu (\citeyear{LeoHsu92}),
Yang and Berger (\citeyear{YanBer94}), Daniels and Kass
(\citeyear{DanKas99}) and Hoff (\citeyear{Hof09}). We present a brief
review of the progress in Bayesian covariance estimation in a somewhat
chronological order starting with priors put on the components of the
spectral decomposition.

\subsection{Priors on the Spectral Decomposition}\label{subsec:Priors}

Starting with the remarkable work of Stein (\citeyear{Ste56,Ste75}),
efforts to improve estimation of a covariance matrix have been confined
mostly to shrinking the eigenvalues of the sample covariance matrix
toward\vadjust{\goodbreak} a common value (Dey and  Srinivasan, \citeyear{DeySri85}; Lin
and Perlman, \citeyear{LinPer85}; Haff, \citeyear{Haf91}; Yang and
Berger, \citeyear{YanBer94}; Daniels and Kass, \citeyear{DanKas99};
Hoff, \citeyear{Hof09}). Such covariance estimators have been shown to
have lower risk than the sample covariance matrix. Intuitively,
shrinking the eigenvectors is expected to further improve or reduce the
estimation risk (Daniels and Kass, \citeyear{DanKas99,DanKas01};
Johnstone and Lu, \citeyear{JohLu09}).

There are three broad classes of priors that are based on unconstrained
parameterizations of a covariance matrix using its spectral
decomposition. These have the goal of shrinking some functions of the
off-diagonal entries of $\Sigma$ or the corresponding correlation
matrix toward a common value like zero. Consequently, estimation of the
$\frac {p(p-1)}{2}$ dependence parameters is reduced to that of
estimating a few parameters.

Perhaps, the first breakthrough with the GLM principles in mind is the
log matrix prior due to Leonard and Hsu (\citeyear{LeoHsu92}) which is
based on the matricial logarithm defined in Section~\ref{spec}. Thus,
formally a multivariate normal prior with a large number of
hyperparameters is introduced. They show the flexibility of this class
of priors for the covariance matrix of a multivariate normal
distribution, yielding much more general hierarchical and empirical
Bayes smoothing and inference, when compared with a conjugate analysis
involving an IW prior. The prior is not conditionally conjugate, and
according to Brown, Le and Zidek (\citeyear{BroLeZid94}), its major
drawback is the lack of statistical interpretability of the entries of
$\log \Sigma$ and their complicated relations to those of $\Sigma$ as
seen in Section~\ref{spec}. Consequently, prior elicitation from
experts  and substantive knowledge cannot be used effectively in
arriving at priors for the entries of $\log \Sigma$ and their
hyperparameters; see Liechty, Liechty and M{\"u}ller
[(\citeyear{LieLieMul04}), page 2] for a discussion on the lack of
intuition and relationship between log-eigenvalues and correlations.

The reference (noninformative) prior for a covariance matrix in Yang
and Berger (\citeyear{YanBer94}) is of the~form
\[
p( \Sigma ) = c \Biggl[| \Sigma | \prod_{i< j} ( \lambda_i
-\lambda_j)\Biggr]^{-1},
\]
where $\lambda_1 > \lambda_2 > \cdots > \lambda_p$ are the ordered
eigenvalues of $\Sigma$ and $c$ is a constant. Yang and Berger
[(\citeyear{YanBer94}), page 1199] note that compared to the Jeffreys
prior, the reference prior puts considerably more mass near the region
of equality of the eigenvalues. Therefore, it is intuitively plausible
that the reference prior would produce a covariance\vadjust{\goodbreak} estimator with
better eigenstructure shrinkage. Furthermore, they point out that the
reference priors for $\Sigma^{-1}$ and the eigenvalues of the
covariance matrix are the same as $p( \Sigma )$. Expression for the
Bayes estimator of the covariance matrix using this prior involves
computation of high-dimensional posterior expectations; the computation
is done using the hit-and-run sampler in a Markov chain Monte Carlo
setup. An alternative noninformative reference prior for $\Sigma$ (and
the precision matrix) which allows for closed-form posterior
estimation is given in Rajaratnam, Massam and Carvalho
(\citeyear{RajMasCar08}).

It is known (Daniels, \citeyear{Dan05}) that the Yang and Berger's
(\citeyear{YanBer94}) reference prior implies a uniform prior on the
orthogonal matrix $P$ and flat improper priors on the logoarithm of the
eigenvalues of the covariance matrix. The shrinkage priors of Daniels
and Kass (\citeyear{DanKas99}) also rely on the spectral decomposition
of the covariance matrix, but are designed to shrink the eigenvectors
by reparametrizing the orthogonal matrix in terms of $\frac
{p(p-1)}{2}$ Givens angles (Golub and Van Loan, \citeyear{GolVan89})
$\theta$ between pairs of the columns of the orthogonal matrix $P$.
Since $\theta$ is restricted to lie in the interval $(-\pi/2,\pi/2)$, a
logit transform will make it unconstrained so as to conform to the GLM
principles. They put a a mean-zero normal prior on the logit
tranformation of the Givens angles. The statistical relevance and
interpretation of the Givens angles are not well understood at this
time. The local parametrization of orthogonal matrices in Boik
(\citeyear{Boi02}) could shed some light on the problem of
interpretation of the new parameters. The idea of introducing matrix
Bingham distributions as priors on the group of orthogonal matrices
(Hoff, \citeyear{Hof09}) could also be useful in shrinking the
eigenvectors of the sample covariance matrix.

Using simulation experiments, Yang and Berger (\citeyear{YanBer94})
compared the performance of their reference prior Bayes covariance
estimator to the covariance estimators of Stein (\citeyear{Ste75}) and
Haff (\citeyear{Haf91}) and found it to be quite competitive based on
the risks corresponding to the loss functions $L_i, i=1,2$. Daniels and
Kass (\citeyear{DanKas99}), also using simulations, compared  the
performance of their shrinkage estimator to several other Bayes
estimators of covariance matrices, using only the risk corresponding to
the $L_1$ loss function. It turns out that the Bayes estimators from
the Yang and Berger's (\citeyear{YanBer94}) reference prior do as well
as those from the Givens-angle prior for some nondiagonal and
ill-conditioned matrices, but suffers when the true matrix is diagonal
and poorly conditioned.\vadjust{\goodbreak}

\subsection{The Generalized Inverse Wishart Priors} \label{ss:CCD}

The use of Cholesky decomposition of a covariance matrix or the
regression dissection of the associated random vector has a long
history and can be traced at least to the work of Bartlett
(\citeyear{Bar33}); see Liu (\citeyear{Liu93}). It is shown by Brown,
Le and Zidek (\citeyear{BroLeZid94}) that a regression dissection of
the inverse Wishart (IW) distribution reveals some of its noteworthy
features, making it possible to define flexible generalized inverted
Wishart (GIW) priors for general covariance matrices.

These priors are constructed by first partitioning a multivariate
normal random vector $Y$ with mean zero and covariance matrix $\Sigma$
into $k \leq p$ subvectors: $Y = (Z_1, \ldots , Z_k)^\prime$, and
writing its joint density as the product of a sequence of conditionals:
\[
f(y) = f(z_1) f(z_2|z_1) \cdots f(z_k | z_{k-1}, \ldots , z_1).
\]
Now, in each conditional distribution one places normal prior
distributions on the regression coefficients and inverse Wishart on the
prediction variances. The hyperparameters can be structured so as to
maintain the conjugacey of the resulting priors. It is known (Daniels
and Pourahmadi, \citeyear{DanPou02}; Rajaratnam, Massam and Carvalho,
\citeyear{RajMasCar08}) that such priors offer considerable
flexibility, as there are many parameters to control the variability in
contrast to the one parameter for IW.

These ideas and techniques have been further refined in Garthwaite and
Al-Awadi (\citeyear{GarAlA01}) in prior distribution elicitation from
experts, and extended to longitudinal and panel data setup in Daniels
and Pourahmadi (\citeyear{DanPou02}) and Smith and Kohn
(\citeyear{SmiKoh02}). The GIW prior was further refined in Daniels and
Pourahmadi (\citeyear{DanPou02}) using the finest partition of $Y$,
that is, using $k=p$. In this case all restrictions on the
hyperparameters are removed from the normal and inverse Wishart (gamma)
distributions and the prior remains conditionally conjugate, in the
sense that the full-conditional of the regression coefficients is
normal given the prediction variances, and the~full-conditional of
prediction variances is inverse gamma given the regression
coefficients. For a review of certain advantages of this approach in
the context of longitudinal data and some examples of analysis of such
data, see Daniels (\citeyear{Dan05}) and Daniels and Hogan
(\citeyear{DanHog08}).

\subsection{Priors on Correlation Matrices}\label{ss:CVCD}

One of the first uses of variance-correlation decom\-position in Bayesian
covariance estimation seems to be due to Barnard, McCulloch and Meng
(\citeyear{BarMcCMen00}),\vadjust{\goodbreak} who, using $p( \Sigma) = p(D,R) =
p(D)p(R|D)$, introdu\-ced independent priors for the standard deviations
in $D$ and the correlations in $R$.

Specifically, they used log normal priors on variances independently of
a prior on the whole matrix~$R$. The latter is capable of inducing
uniform $(-1,1)$ priors on the entries $\rho_{ij}$ of the
correlation~ma\-trix~$R$; see Liu and Daniels (\citeyear{LiuDan06}). This is done by
first deriving the marginal distribution of~$R$ when~$\Sigma$ has a
standard IW distribution, $W^{-1}_p (I, \nu ), \nu \geq p$, with the
density
\[
f_p (\Sigma | \nu ) = c | \Sigma|^{- (1/2) (\nu + p + 1)} \exp \bigl( -
\tfrac{1}{2} \operatorname{tr} \Sigma^{-1} \bigr).
\]
It turns out that
\[
f_p (R| \nu ) = c|R|^{(1/2) (\nu - 1)(p-1)-1} \prod^p_{i=1} |
R_{ii}|^{- \nu / 2},
\]
where $R_{ii}$ is the principal submatrix of $R$, obtained by deleting
its $i$th row and column. Then, using the marginalization property of
the IW (i.e., a principal submatrix of an IW is still an IW), the
marginal distribution of each $\rho_{ij}$, $i \neq j$, is obtained as
\[
f( \rho_{ij} | \nu ) = c(1 - \rho^2_{ij})^{(\nu - p - 1)/2}
,\quad |\rho_{ij} | \leq 1.
\]
The latter can be viewed as a Beta $ ( \frac{\nu - p + 1}{2} ,
\frac{\nu - p + 1}{2} )$ on $(-1,1)$, which is uniform when $\nu = p +
1$. Moreover, by choosing $p \leq \nu < p+1$ or $\nu > p+1$, one can
control the tail of $f( \rho_{ij} | \nu )$, that is, making it heavier
or lighter than the uniform. Thus, the above family of priors for $R$
is indexed by a single ``tuning'' parameter $\nu$.

Liechty, Liechty and M{\"u}ller (\citeyear{LieLieMul04}) note that few
existing probability models and parameterizations for covariance
matrices allow for easy interpretation and prior elicitation. They
propose priors in which correlations are grouped based on similarities
among the correlations or based on groups of variables. A~good example
of this situation is in financial time series where it is often known
that returns of stocks in the same industries are more closely related
than others.


\subsection{Reparameterization via Partial Autocorrelations} \label{subsec:pacf}

In this section we present yet another unconstrai\-ned and statistically
interpretable reparameterization of~$\Sigma$, but now using the notion
of partial autocorrelation function (PACF) from time series analysis
(Box, Jenkins and Reinsel, \citeyear{BoxJenRei94}; Pourahmadi,
\citeyear{Pou01}, Chapter 7). As expected, this approach, just like the
Cholesky decomposition, requires an \textit{a priori} order among the
random variables in $Y$. It is motivated\vadjust{\goodbreak} by and tries to mimic the
phenomenal success of the PACF of a stationary time series in model
formulation (Box, Jenkins and Reinsel, \citeyear{BoxJenRei94}) and
removing the positive-definiteness constraint on the autocorrelation
function (Ramsey, \citeyear{Ram74}). We note that reparameterizing the
stationarity-invertibility domain of ARMA models by Jones
(\citeyear{Jon80}) had a~profound impact on algorithms for computing
the MLE of the ARMA coefficients  and guaranteeing that the estimates
are in the feasible region.

Starting with the variance-correlation decomposition, we focus on
reparameterizing the correlation matrix $R = ( \rho_{ij})$ in terms of
entries of a simpler symmetric matrix $\Pi = ( \pi_{ij})$, where
$\pi_{ii} = 1$ and for \mbox{$i < j$}, $\pi_{ij}$ is the \textit{partial
autocorrelation} between $y_i$ and $y_j$ adjusted for the
\textit{intervening} (not the remaining) variables. More precisely,
$\pi_{i,i+1} = \rho_{i,i+1}, i = 1, \ldots,\allowbreak p-1$, are the lag-1
correlations and for \mbox{$j - i \geq 2$}, $\pi_{ij} = \rho_{ij | i+1, \ldots ,
j-1}$ in the notation of Anderson (\citeyear{And03}), page~41. Note
that, unlike $R$, and the matrix of full partial correlations
$(\rho^{ij})$ constructed from $\Sigma^{-1}$ in
Section~\ref{sec:precision}, $\Pi$ has a much simpler structure in that
its entries are free to vary in the interval $(-1,1)$. If needed, using
the Fisher $z$-transform $\Pi$ can be mapped to the matrix
$\tilde\Pi$ where its off-diagonal entries take values in
the entire real line $(- \infty , + \infty )$.

Compared to the long history of using the PACF in time series analysis
(Quenouille, \citeyear{Que49}), research on establishing a one-to-one
correspondence between a general covariance matrix and $(D, \Pi )$  has
a rather short history. An early work in the Bayesian context is due to
Eaves and Chang (\citeyear{EavCha92}), followed by Zimmerman
(\citeyear{Zim00}) and Pourahmadi [(\citeyear{Pou99,Pou01}), pa\-ge~102]
for longitudinal data, D\'{e}gerine and Lambert-Lacroix
(\citeyear{DegLam03}) for the nonstationary time series, and Kurowicka
and Cooke (\citeyear{KurCoo03}) and Joe (\citeyear{Joe06})~for
a~general random vector. The fundamental determi\-nantal identity,
\begin{equation} \label{eq:id}
| \Sigma | = \Biggl( \prod^p_{i=1} \sigma_{ii} \Biggr) \prod^p_{i=2}
\prod^{i-j}_{j = 1} ( 1 - \pi^2_{ij} ) ,
\end{equation}
has been redicovered recently by D\'{e}gerine and Lam\-bert-Lacroix
(\citeyear{DegLam03}), Kurowicka and Cooke (\citeyear{KurCoo03}) and
Joe (\citeyear{Joe06}), but its origin can be traced to a notable and
somewhat neglected paper of Yule (\citeyear{Yul07}), equation (25).

The identity (\ref{eq:id}) plays a central role in Joe's
(\citeyear{Joe06}) method of generating random correlation matrices
whose distributions are \textit{independent of the order of variables}
in $Y$. It is used in Daniels and Pourahmadi (\citeyear{DanPou09}) to
introduce priors for the Bayesian analysis of correlation matrices.
These papers  employ\vadjust{\goodbreak} independent linearly transformed Beta priors on
$(-1,1)$ for the partial autocorrelations $\pi_{ij}$. However, Jones
(\citeyear{Jon87}) seems to be the first to use such Beta priors in
simulating data from ``typical'' ARMA models.

\section{Conclusions} \label{sec:what}

We have reviewed progress in covariance estimation for low- and
high-dimensional data, from the narrow perspectives of the GLM and
regularization or parsimony and sparsity. Recent appearance of many
regression-based techniques and the use of LASSO-type penalties show
that covariance estimation can benefit greatly from mimicking/using the
conceptual and computational tools of regression ana\-lysis. Fortunately,
mostly due to the computational-algorithmic advances centered around
LASSO, the high-dimensionality challenge in covariance estimation has
been become manageable, however, the posi\-tive-definiteness challenge
still remains. Its removal could not only further reduce the
computational cost due to high-dimensionality, but is also crucial
for~par\-simony and writing simple, interpretable models using covariates.
Among the three \mbox{matrix} decompositions, the spectral and Cholesky
decompositions are the most helpful in removing the
positive-definiteness constraint. These along with some recent
covariance estimation algorithms enforcing the positive-defini\-teness
suggest that there are trade-offs among the requirements of
unconstrained parameterization, statistical interpretability and the
computational cost.

In summary, the problem of removing the positive-definiteness
constraint remains open, in the sense that, as yet, no
\textit{unconstrained and statistically interpretable}
reparameterization exists for a general covariance matrix without
imposing an order on the variables.

\section*{Acknowledgments}

Research supported in part by the NSF Grants DMS-05-05696 and
DMS-09-06252. Comments from the Associate Editor and two referees have
greatly improved the presentation, focus and scope of the paper.


\end{document}